\documentclass[conference]{IEEEtran}

\usepackage{graphicx}
\usepackage[sort&compress,numbers]{natbib}
\usepackage{amssymb}
\usepackage{amsmath}
\usepackage{amsfonts}
\usepackage{colortbl}
\usepackage{hhline}
\usepackage{color}
\usepackage[bf,small]{caption}
\usepackage{indentfirst}
\usepackage{url}
\usepackage{subfigure}
\usepackage{tikz}
\usepackage{bbding}
\usepackage[all]{xy}
\usepackage{array}
\usepackage{multirow}
\usepackage{multicol}
\usepackage{balance}
\usepackage{etex}
\usepackage{grffile}
\usepackage{paralist}
\usepackage[ruled,noend]{algorithm2e}

\newcommand{\descr}[1]{\vspace{0.25cm} \noindent \textbf{#1}}
\usepackage{url}
\usepackage{color}
\definecolor{darkblue}{RGB}{0,0,120}
\usepackage[bookmarks=false, colorlinks=true, plainpages=false,  linkcolor=darkblue,   citecolor=darkblue, urlcolor=darkblue, filecolor=darkblue]{hyperref}
\usepackage{breakurl}

\begin{document}

\title{A Comparative Usability Study of\\ Two-Factor Authentication\thanks{A preliminary version of this paper appears in USEC 2014.}}

\author{\IEEEauthorblockN{Emiliano De Cristofaro$^\dag$}
\IEEEauthorblockA{University College London\\
e.decristofaro@ucl.ac.uk}
\and
\IEEEauthorblockN{Honglu Du}
\IEEEauthorblockA{PARC\\
honglu.du@parc.com}\\
\and
\IEEEauthorblockN{Julien Freudiger}
\IEEEauthorblockA{PARC\\
julien.freudiger@parc.com}\\
\and
\IEEEauthorblockN{Greg Norcie$^\dag$}
\IEEEauthorblockA{Indiana University\\
gnorcie@indiana.edu}
}

\maketitle

\begin{abstract}
Two-factor authentication (2F) aims to enhance resilience of password-based authentication by requiring users to provide an additional authentication factor, e.g., a code generated by a security token. However, it also introduces non-negligible costs for service providers and requires users to carry out additional actions during the authentication process.
In this paper,  we present an exploratory comparative study of the usability of 2F technologies. First, we conduct a pre-study interview to identify popular technologies as well as contexts and motivations in which they are used. We then present the results of a quantitative study based on a survey completed by 219 Mechanical Turk users, aiming to measure the usability of three popular 2F solutions: codes generated by security tokens, one-time PINs received via email or SMS, and dedicated smartphone apps (e.g., Google Authenticator). We record contexts and motivations, and study their impact on perceived usability.

We find that 2F technologies are overall perceived as usable, regardless of motivation and/or context of use. We also present an exploratory factor analysis, highlighting that three metrics -- ease-of-use, required cognitive efforts, and trustworthiness -- are enough to capture key factors affecting 2F usability.
\end{abstract}

\renewcommand{\thefootnote}{\fnsymbol{footnote}}
\footnotetext{$^\dag$Work done while authors were at PARC.}

\section{Introduction}

Despite their well-known security issues, passwords are still the most popular method of end-user authentication. Guessing and offline dictionary attacks on user-generated passwords are often possible due to their limited entropy. According to Ashlee Vance~\cite{hackme}, 20\% of passwords are covered by a list of only 5,000 passwords. Therefore, aiming to increase security, complex password policies are often enforced, by requiring, e.g., a minimum number of characters,  inclusion of non alpha-numeric symbols, or frequent password expiration. This often creates an undesirable conflict between security and usability -- as highlighted in the context of password selection~\cite{egelman2013does}, management~\cite{karole2011comparative,weiss2008passshapes} and composition~\cite{komanduri2011passwords,von2013survival} -- and drives users to find the easiest password that is policy-compliant~\cite{adams1999users}.
Multi-factor authentication has emerged as an alternative way to improve security by requiring the user to provide more than one authentication {\em factor}, as opposed to only a password. Authentication factors are usually of three kinds:
\begin{enumerate}
\item {\em Knowledge} -- something the user knows, e.g., a password;
\item {\em Possession} -- something the user has, e.g., a security token (also known as hardware token);
\item {\em Inherence} -- something the user is, e.g., a biometric characteristic. 
\end{enumerate}

\renewcommand{\thefootnote}{\arabic{footnote}}

In this paper, we concentrate on the most common instantiation of multi-factor authentication, i.e., the one based on two factors, which we denote as 2F.
Historically, 2F has been deployed mostly in enterprise, government, and financial sectors, where sensitivity of information and services has driven institutions to accept increased implementation/maintenance costs, and/or to impose additional actions on authenticating users. In 2005, the United States' Federal Financial Institutions Examination Council officially recommended the use of multi-factor authentication~\cite{council2005}, thus pressuring most institutions to adopt some forms of 2F authentication for online banking. Similarly, government agencies and enterprises often require employees to use 2F for, e.g., VPN authentication or B2B transactions. More recently, an increasing number of service providers, such as Google, Facebook, Dropbox, Twitter, GitHub, have also begun to provide their users with the option of enabling 2F, arguably, motivated by the increasing number of password databases hacked. Recent highly publicized incidents affected, among others, Dropbox, Twitter, Linkedin, Rockyou.

Alas, security of 2F also suffers from a few limitations: 2F technologies, including recently proposed ones based on fingerprints~\cite{iphone}, are often vulnerable to man-in-the-middle, forgery, or Trojan-based attacks, and are not completely effective against phishing~\cite{schneier2005two}. Furthermore, 2F systems introduce non-negligible costs for service providers and require users to carry out additional actions in order to authenticate, e.g., entering a one-time code and/or carrying an additional device with them. A common assumption in the IT sector, partially supported by prior work~\cite{bauer2007lessons,bonneau2012quest,braz2006security,gunson2011user,sabzevar2008universal,strouble2009productivity}, is that 2F technologies have low(er) usability compared to authentication based only on passwords, and this likely hinders larger adoption. 
In fact, a few start-up companies (e.g., PassBan, Duo Security, Authy, Encap) aim to innovate the 2F landscape by introducing more usable solutions to the market.
However, little research actually studied the usability of different 2F technologies. %

We begin to address this gap by presenting an exploratory comparative usability study. First, we conduct a pre-study interview, involving 9 participants, to identify popular 2F technologies as well as the contexts and motivations in which they are used. Then, we present the design and the results of a quantitative study (based on a survey involving 219 Mechanical Turk users), that aims to
measure the usability of a few second-factor solutions: one-time codes generated by security tokens, one-time pins received via SMS or email, and dedicated smartphone apps (such as, Google Authenticator).
Note that all our participants make use of 2F (i.e., had been forced to or had chosen to), and thus might already have a reasonable mental model of how 2F works.

Our comparative analysis of 2F usability yields some interesting findings. We show how users' perception of 2F usability is often correlated with their individual characteristics (such as, age, gender, background), rather than with the actual technology or the context in which it is used.
We find that, overall, 2F technologies are perceived as highly usable, with little difference among them, not even when they are used for different motivations and different contexts. 
We also present an exploratory {\em factor analysis}, which demonstrates that three metrics -- ease-of-use, required cognitive efforts, and trustworthiness -- are enough to capture key factors affecting the usability of 2F technologies.
Finally, we pave the way for future qualitative studies, based on our factor analysis, to further analyze our findings and confirm their generalizability.

\section{Related Work}\label{sec:rw}
In this section, we review prior work on the usability of single- and multi-factor authentication technologies.

\subsection{Usability of Single Factor Technologies}
Adams and Sasse~\cite{adams1999users} showed that, for users, security is not a primary task, thus users feel under attack by ``capricious'' password polices. Password policies often mandate the use of long (and hard-to-remember) passwords, frequent password changes, and using different passwords across different services. This ultimately drives the user to find the simplest password that barely complies with requirements~\cite{adams1999users}. Inglesant and Sasse~\cite{inglesant2010true} analyzed ``password diaries'', i.e., they asked users to record the times they authenticated via passwords, and found that frequent password changes are a burden, users do not change passwords unless forced to, and that it is difficult for them to create memorable, secure passwords adhering to the policy. They also concluded that context of use has a significant impact on the ability of users to become familiar with complex passwords and, essentially, on their usability.

Bardram et al.~\cite{bardram2005trouble} discussed burdens on nursing staff created by hard-to-remember passwords in conjunction with frequent logouts required by healthcare security standards, such as the Health Insurance Portability and Accountability Act (HIPAA).
The impact on usability and security of password composition policies has also been studied. For instance, Komanduri et al.~\cite{komanduri2011passwords} showed that complex password policies can actually \textit{decrease} average password entropy, and that a 16-character with no additional requirements provided the highest average entropy per password. 
Egelman et al.~\cite{egelman2013does} found that for
``important'' accounts, a password meter (i.e., a visual clue on password's strength)  successfully helps increase entropy.  

Another line of work has focused on {\em password managers}.
Chiasson et al.~\cite{chiasson2006usability} compared the usability of two password managers (PwdHash and Password Multiplier), pointing to a few usability issues in both implementations and showing that users were often uncomfortable ``relinquishing control'' to password managers. 
Karole et al.~\cite{karole2011comparative} studied the usability of three password managers (LastPass, KeePassMobile, and Roboform2Go), with a focus on mobile phone users. They concluded that users preferred portable, stand-alone managers over cloud-based ones, despite the better usability of the latter, as they were not comfortable giving control of their passwords to an online entity. 

Finally, Bonneau et al.~\cite{bonneau2012quest} evaluated, without conducting any user study, authentication schemes including: plain passwords, OpenID~\cite{recordon2006openid}, security tokens, phone-based tokens, etc. They used a set of 25 subjective factors: 8 measuring usability, 6 measuring deployability, and 11 measuring security. Although they did not conduct any user study, authors concluded that: (i) no existing authentication scheme does best in all metrics, and (ii) technologies that one could classify as 2F do better than passwords in security but worse in usability.

Although not directly related to our 2F study, we will use in our factor analysis some metrics introduced in the context of password replacements~\cite{bonneau2012quest} and password managers~\cite{karole2011comparative}. 

\subsection{Usability of Multi-Factor Authentication Technologies}

Previous work has suggested that security via 2F decreases usability of end-user authentication. For instance, Braz et al.~\cite{braz2006security} showed that 2F increases ``redundancy'', thus augmenting security but decreasing usability. Along similar lines, Strouble et al.~\cite{strouble2009productivity} analyzed the effects of implementing 2F on productivity, focusing on the ``Common Access Card'' (CaC), a combined smart card/photo ID card used (at that time) by US Department of Defense (DoD) employees. They reported that users stopped checking emails at home (due to the unavailability of card readers) and that many employees accidentally left their card in the reader. Authors also estimated that the DoD spent about \$10.4M on time lost (e.g., when employees left the base without their card and were unable to re-enter) and concluded that the CaC increased security at the expense of productivity.
 
Gunson et al.~\cite{gunson2011user} focused on the usability of single and two-factor authentication in automated telephone banking. They presented a survey involving $62$ users of telephone banking, where participants were asked to rate their experience using a proposed set of $22$ usability-related questions. According to their analysis, 2F was perceived to be more secure, but again less usable, than simple passwords and PINs.

Weir et al.~\cite{weir2009user} compared usability of push-button tokens, card activated tokens, and PIN activated tokens. They measured usability in terms of efficiency (time needed for authentication), 
as well as in terms of satisfaction, by asking users to rate their experience using a  set of $30$ questions. In addition to usability, 
they measured quality, convenience, and security. They showed that users value convenience and usability over security, and thus quality and usability are sacrificed when increasing layers of security are required. 

Somewhat closer to our work is another study by Weir et al.~\cite{weir2010usable}, which analyzed the usability of passwords and two methods of 2F: codes generated by token and PINs received via SMS. They performed a lab study where $141$ participants were asked to report on the usability of the three technologies using $30$ proposed questions. The authors concluded that familiarity with a technology (rather than perceived usability) impacted user willingness to use a given authentication technology. Their results showed that users perceived the 1-factor method (with which the average user had most experience) as being the most secure and most convenient option. 

Our work differs from Weir et al. in several key aspects. We compare a larger diversity of 2F technologies (security tokens, codes received via SMS/email, and dedicated apps) and do not study the trade-off between security and usability. In constrast, we provide a comparative study among different technologies, aiming to understand how each 2F technology performs compared to others. Specifically, we study the relation between 2F technologies and the contexts in which they are used, as well as the motivation driving the users to adopt them, partially motivated by previous work by Goffman~\cite{goffman1959presentation} and Nissenbaum~\cite{nissenbaum2004privacy}, who showed that human behavior often significantly differs based on context. 
Finally, we consider a larger pool of participants, measure an extensive list of factors beyond Weir's work, and conduct an exploratory factor analysis to determine key factors that affect usability of 2F.

\section{Pre-Study Interviews with 2F Users}
Our first step is to determine broad trends and attitudes of 2F users: we aim to obtain a 
a general understanding of 2F technologies in use, the context in which these technologies are deployed, and why they are adopted. To this end, we interviewed 9 participants, and designed a larger quantitative study (detailed in next section) based on them.
Both user studies were approved by PARC's Institutional Review Board.

\subsection{Methodology}

We recruited participants by posting to local mailing lists
and social media (Google+ and Facebook), announcing paid interviews 
for a user study on security and authentication technologies.
Interested users were invited to complete an online screening survey to assess eligibility to participate. We collected basic demographic information such as age, gender, 
education level, familiarity with Computer Security, and asked potential participants whether or not they had previously used 2F. Users without 2F familiarity were not invited to participate. 

The screening survey was completed by 29 people, and we selected 9 participants, most of them from the Silicon Valley, with a wide range of ages (21 to 49), genders (5 men, 4 women), and educational background (ranging from high school to Ph.D. degrees). 5/9 users reported having a background in Computer Security.

We interviewed users in one-on-one meetings, either face to face, or via Skype. Before each interview, users were given a consent form, indicating the interview procedure and data confidentiality. Each participant was compensated with a \$10 Amazon Gift Card.

We started the interviews by reading from a list of 2F technologies, asking participants if they had used them: 
\begin{compactitem}
\item PIN from a paper/card (one-time PIN)
\item A digital certificate
\item An RSA token code
\item A Verisign token code
\item A Paypal token code
\item Google Authenticator
\item A PIN received by SMS/email
\item A USB token
\item A smartcard
\end{compactitem}

To assess users' understanding and familiarity with 2F, we let them provide a brief 
description of two-factor authentication, and explain the difference with password-based authentication. (Obviously, we did not provide users with a 2F definition prior to this question, nor mention that the study was about 2F).

Then, we asked participants {\em why} they used 2F and why they thought other people would; this helped us understand the motivation and the context in which they used 2F.
Users were also asked to recall the last time they had used any 2F technology and report any encountered issues and whether or not they wanted to change the technology (and, if so, how). If users had used multiple technologies, we also asked to compare them, and this helped us understand how participants use and perceive 2F technologies.

\subsection{Findings}
We found that most used 2F technologies included: codes generated by a \emph{security token}, received via \emph{SMS or email}, and codes generated by a dedicated \emph{smartphone app}, entered along with username and password.

Participants used 2F technology in three contexts: \emph{work} (e.g., to log into their company's VPN), \emph{personal} (e.g., to protect a social networking account), or \emph{financial} (e.g., to gain access to online banking).

Study participants used 2F because they either were \emph{forced}, \emph{wanted to}, or \emph{had an incentive}. Most users adopted security tokens because an employer or bank had forced them. Some were unhappy about this: A participant mentioned 2F was not ``worth spending 5 minutes for \$1.99 purchases''. Two participants (customers of different banks) reported adopting 2F in order to ``obtain higher limits on online banking transactions.'' Other users used 2F to ``avoid getting hacked.'' 

Some users of tokens complained that it was annoying to have to remember to carry security tokens. One user recommended to ``store the token in the laptop bag'' to avoid this issue. Some users experienced delays from SMS-based codes, and
were ``annoyed, especially when paying for incoming texts.'' One user pointed out that (s)he ``preferred text messages'', since
(s)he ``did not have a smartphone.'' Others preferred not to use security tokens as they ``can be lost.'' Some participants
preferred tokens as they are easier to use compared to mobile applications, where one has to ``look down to unlock screen, find
app, open app, and read the code.''

\section{Quantitative Analysis of 2F Users' Preferences}\vspace{0.2cm}

Our second/main study consists of a quantitative analysis of 2F users' preferences.
Inspired by the results of our pre-study interviews, we designed and conducted a survey involving 219 2F users, recruited on Mechanical Turk (MTurk). 

\subsection{Methodology} 
We initially recruited 268 U.S.-based MTurk users. All MTurk users had to have a 95\% or higher approval rating. 13 of them were not eligible as they had not used any 2F technology. 36 users abandoned the survey prior to completion. The remaining 219 MTurk users were asked to complete an online survey about 2F technologies. Study participants received \$2.00 for no more than 30 minutes of survey taking. 

\descr{On MTurk studies.} Previous research showed that MTurk users are a valid alternative to traditional human subject pools in social sciences. For example, Jakobsson~\cite{jakobsson2009experimenting} compared the results of a study conducted on MTurk with results conducted by an independent survey company, and found that both results were statistically indistinguishable. Furthermore, MTurk users are often more diverse in terms of age, income, education level, and geographic location than the traditional pool for social science experiments~\cite{henrich2010weirdest}.
However, research has also highlighted that MTurkers are often younger and more computer savvy~\cite{Christenson:2013}. As we will discuss in Sec.~\ref{sec:discussion}, our work is intended to serve as a preliminary study, which should guide and inform the design of a qualitative study.

\descr{Data sanitization.} Kittur et al.~\cite{kittur2008crowdsourcing} point out that MTurk users often try to cheat at tasks. Therefore, we designed the survey to include several sanity-check questions, such as simple math questions (in the form of Likert questions) in order to verify that participants were paying attention (and were not answering randomly). We also introduced some contrasting Likert questions (e.g., ``I enjoyed using the technology'' and ``I did not enjoy using the technology'') and verified that answers were consistent. Users who did not answer correctly all sanity checks were to be discarded from the analysis (but still compensated). Actually, all users answered the sanity-checks correctly. 
Also note that analysis of the time spent by each survey participant showed completion times in line with those of test runs done by experimenters (~15--30 minutes, depending on number of 2F tech used). %

\descr{Recruitment.} We screened potential participants by asking whether they had used 2F, and presented a list of examples: security tokens, codes received via SMS/email, and dedicated smartphone apps. Users who reported to have never used any of these technologies were told that they were not eligible to participate in our survey, and blocked from proceeding further or going back to change their answer. %
Also note the MTurk task announcement did not state that users were required to have used 2F and merely presented it alongside other basic demographics such as age and gender. 

\descr{Demographics.} The demographics of the 219 study participants are reported in Table~\ref{demo}. Our population included 135 (61.6\%) males and 84 females (38.4\%). 50/219 (22.8\%) users reported a background in computer science, and 12/219 (5.4\%) users reported a background in computer security. Education levels ranged from high school diploma to PhD degrees. Ages ranged from 18 to 66, with an average age of 32 and a standard deviation of 10.2.

\begin{table}[ttt]
\centering
\begin{tabular}{|l r|}
\hline
\textbf{Gender} & \\
\hline
Male & 61.6\% \\
Female & 38.4\% \\
\hline \hline
\textbf{Age}& \\
\hline
18--24 & 22.4 \%\\
25--34 & 48.4 \%\\
35--44 & 17.8 \%\\
45--54 & 5.4 \%\\
55--65 & 5.4 \%\\
65+ & 0.5 \%\\
\hline \hline
\textbf{Income}& \\
\hline
Less than \$10,000 & 15.5 \%\\
\$10,000 -- \$20,00 & 14.6 \%\\
\$20,001 -- \$35,000 & 25.5\%) \\
\$35,001 -- \$50,000 & 18.3 \%\\
\$50,0001-- \$75,000 & 18.7 \%\\
\$75,001 -- \$90,000 & 3.6 \%\\
\$90,0001 -- \$120,000 & 2.7 \%\\
\$120,001 -- \$200,000 & 0.9 \%\\
\hline \hline
\textbf{Education}& \\
\hline
Less than high school & 0.46 \%\\
Some college & 32 \%\\
Undergrad & 37.4 \%\\
Some grad school & 3.1\% \\
Master's degree & 5.9 \%\\
PhD & 0.9 \%\\
\hline \hline
\textbf{Familiar with Computer Science?} & \\
\hline
Yes & 22.8\%\\
No & 77.2\%\\
\hline
\hline
\textbf{Familiar with Computer Security?}& \\
\hline
Yes & 5.4\%\\
No & 94.6\%\\
\hline
\end{tabular}
\vspace{0.2cm}
\caption{Participants' demographics (Total n = 219).\label{demo}}
\vspace{0.1cm}
\end{table}

\begin{figure*}[ttt]
\centering
\subfigure[]{
  \includegraphics[width=.295\linewidth]{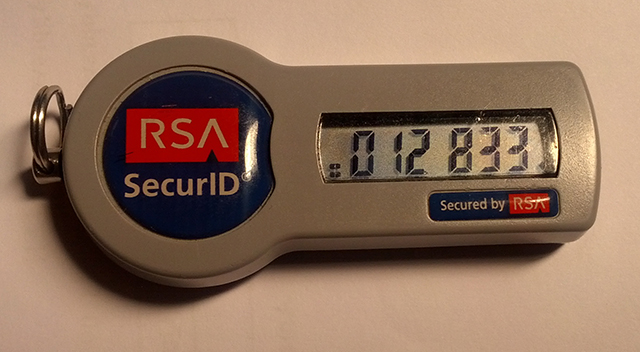}
  \label{fig:token} 
}
\subfigure[]{
  \includegraphics[width=.375\linewidth]{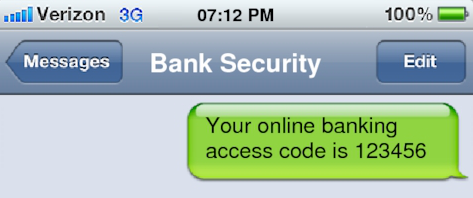}
  \label{fig:text} 
}
\subfigure[]{%
  \includegraphics[scale=.47]{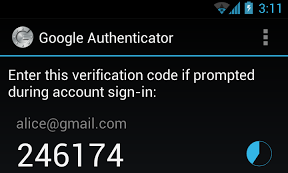}
  \label{fig:app} 
}
\caption{Examples of 2F technologies: (a) codes generated by a security token, (b) codes received via SMS, (c) codes generated by a dedicated smartphone app.}
\vspace{0.1cm}
\end{figure*}

\subsection{Study Design} %

\noindent\textbf{Technologies, Context, and Motivation.} The first question in the survey asked users if they had used any of the following 2F technologies (for each of them, we displayed an example picture):

\begin{compactitem}
\item {\em Token:} Standalone pieces of hardware which display a code, Figure \ref{fig:token}.
\item {\em Email/SMS:} A code received via email or SMS (also known as ``text messages''), Figure \ref{fig:text}.
\item {\em App:} Codes delivered via an app running on a smartphone or other portable electronic device, such as an iPad or Android tablet, Figure \ref{fig:app}.
\end{compactitem}

Next, the survey branched depending on how many and what technology/technologies had been selected.
Specifically, users were asked to answer the same set of questions for each technology they had used.

One of our main objectives was to measure and compare in which context and with what motivation users were exposed to 2F technologies. Specifically, for each technology we asked users in which of the following {\em context(s)} they used the technology:
\begin{compactitem}
\item {\em Financial:} While doing online banking or other financial transactions (e.g., bill payment, checking credit card
balance, doing taxes).
\item {\em Work:} While performing work duties (e.g. logging in company VPN).
\item {\em Personal:} While accessing a personal account not used for work or finance (e.g., Facebook, Twitter, Google, etc.).
\item {\em Other:} Open-ended.
\end{compactitem}

Also, we asked users {\em why} they had been using 2F. Possible motivations included:
\begin{compactitem}
\item {\em Voluntary:} The participant voluntarily adopted 2F .
\item {\em Incentive:} The participant got an incentive to adopt 2F (e.g., extra privileges/functionality, such as increased bank transfer limits).
\item {\em Forced:} The participant had no choice (e.g., employer policy forcing adoption).
\item {\em Other:} Open-ended.
\end{compactitem}

\descr{System Usability Score and Other Likert Questions.}
For each employed 2F technology participants were asked to rank the usability of the technology using $10$ Likert questions from the System Usability Scale (SUS)~\cite{brooke1996sus}. Previous research has shown SUS is a fairly accurate measure of usability~\cite{bangor2008empirical}.

Note that, in order to be consistent with other Likert questions in our survey, we modified the SUS questionnaire to include a 7-point range, rather than the more common 5-point range, with 1 being ``Strongly Disagree'' and 7 being ``Strongly Agree''. %

Next, for each employed 2F technology, participants, where asked a series of 7-point Likert questions (with 1 being ``Strongly Disagree'' and 7 being ``Strongly Agree'') about the following statements:

\begin{compactitem}
\item \textsf{\small Convenient}: I thought (technology) was convenient.
\item \textsf{\small Quick}: Using (technology) was quick.
\item \textsf{\small Enjoy}: I enjoyed using (technology). 
\item \textsf{\small Reuse}: I would be happy to use (technology) again. 
\item \textsf{\small Helpful}: I found using (technology) helpful. 
\item \textsf{\small No Enjoy}: I did not enjoy using (technology). 
\item \textsf{\small User Friendly}: I found (technology) technology user friendly. 
\item \textsf{\small Need Instructions}: I needed instructions to use (technology). 
\item \textsf{\small Concentrate}: I had to concentrate when using (technology). 
\item \textsf{\small Stressful}: Using (technology) was stressful. 
\item \textsf{\small Match}: (technology) did not match my expectations regarding the steps I had to follow to use it. 
\item \textsf{\small Frustrating}: Using (technology) was frustrating. 
\item \textsf{\small Trust}: I found using (technology) trustworthy. 
\item \textsf{\small Secure}: How secure did you feel to authenticate using (technology) instead of just username \& password? (1: ``Not at All Secure'' 7: ``Very secure'')
\item \textsf{\small Easy}: Knowing how to get the code from (technology) was easy. 
\end{compactitem}

The above questions are inspired from metrics used in previous work~\cite{bonneau2012quest,karole2011comparative} and findings from our pre-study interviews. They are meant to be extensive and measure factors beyond the System Usability Score, such as trustworthiness, convenience, ease of use, reuse, enjoyment, concentration, portability, etc.

\subsection{Results}
We first analyze how 2F technologies are used by investigating the relation between independent factors such as context, motivation, technologies, and gender. We then provide an exploratory factor analysis about users' perception of 2F technologies (Likert questions), aiming to understand which factors are best to capture the usability of 2F. We then provide a comparative analysis of the usability of 2F technologies using those factors, and conclude with a discussion of our findings and highlighting some issues with 2F. 

\begin{table}
\begin{center}
\small
\begin{tabular}{| c | c | c |}
  \hline                        
  \textbf{Group} & \textbf{2F Technologies} & \textbf{\# of Participants} \\
  \hline    
  1 & Token & 11 \\\hline  
  2 & Email/SMS & 77 \\\hline  
  3 & App & 7 \\\hline  
  4 & Token \& Email/SMS & 29 \\\hline  
  5 & Token \& App & 3 \\\hline  
  6 & Email/SMS \& App & 50 \\\hline  
  7 & Token, Email/SMS \& App & 41 \\\hline  
  \textbf{Total} & & \textbf{219}\\
    \hline
\end{tabular}
\end{center}
\vspace{0.2cm}
\caption{Usage of 2F technologies among survey participants.\label{tab:usage}}
\vspace{-0.2cm}
 \end{table}

\descr{Use of 2-Factor.}
Recall from our study design that participants were asked to identify the different 2F technologies they use, in which context, and why. Almost half of the participants ($43\%$) used only one technology, while $37\%$ used two, and $20\%$ three technologies.
Table~\ref{tab:usage} summarizes the use of the three 2F technologies among the $219$ participants. We observe that ``Email/SMS'' (i.e., one-time codes received via SMS or email) is the most used technology as $89.95\%$ ($197/219$) used it as a second factor. Also, $45.20\%$ ($99/219$) of participants used ``App'' (i.e., codes generated by a dedicated smartphone app, such as Google Authenticator). ``Token'' (i.e., codes generated by a hardware/security token) is the least common technology, only used by $24.20\%$ ($53/219$). 
It is interesting to observe that App, despite being the most recent technology, has a higher adoption rate than Token, one of the oldest technology. This evolution might be related with the fast-increasing number of users owning smartphones, which can serve as a second-factor device that is always with the user.

\begin{figure}
\centering
  \includegraphics[scale=.7]{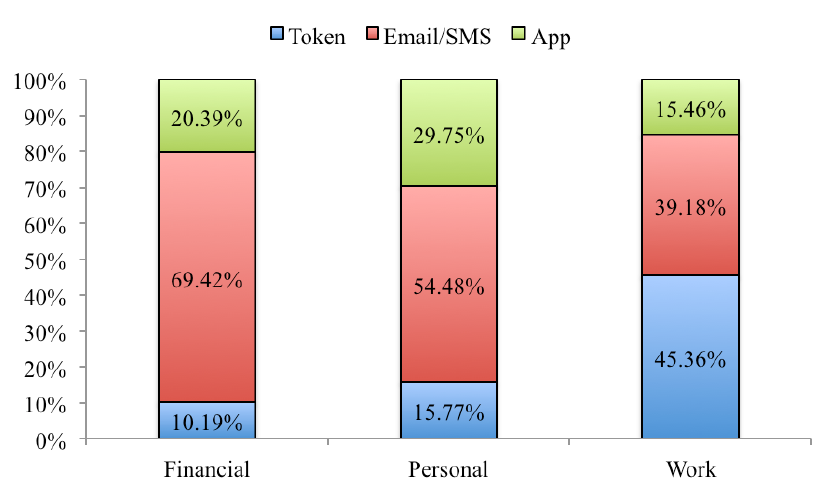}
  \caption{Distribution of the use of 2F technologies across contexts.}
  \label{fig:technologyvscontext} 
\end{figure}

\descr{Different Technologies in Different Contexts.}
The three 2F technologies are used differently depending on context (Figure~\ref{fig:technologyvscontext}). In the financial context, Email/SMS is the most popular 2F ($69.42\%$), followed by App ($20.39\%$) and Token ($10.19\%$). In the personal context, Email/SMS is also the most popular ($54.48\%$), followed by App ($29.75\%$) and Token ($15.77\%$). In the work context, Token is the most popular ($45.36\%$), followed by Email/SMS ($39.18\%$) and App ($15.46\%$). A $\chi^2$-test shows that differences are significant ($\chi^2(4, N=582)=65.18$, $p<0.0001$). No participant reported using 2F in any context other than work/financial/personal (participants could do so via an open-ended question).

It is relatively unsurprising that Token is most popular in the work context---an environment with high inertia---while it is noticeable that many users adopt tokens in the personal context.
The analysis of open-ended questions seem to show that online gaming is the main field of adoption for Token in the personal context.

\begin{figure}
\centering
  \includegraphics[scale=.7]{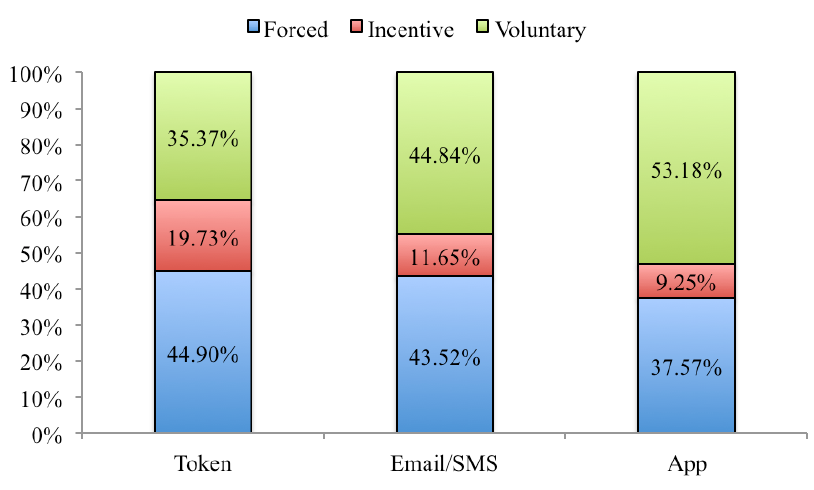}
  \caption{Distribution of the motivation across 2F technologies.}
  \label{fig:technologyvsmotivation} 
\end{figure}

\descr{Different Motivations for Different Technologies.}
We find that few participants are incentivized to use 2F -- see Figure~\ref{fig:technologyvsmotivation}.  Only $19.73\%$ of Token users, $11.65\%$ of Email/SMS users and $9.25\%$ of App users are incentivized. Actually, $44.90\%$ of Token users were forced, while $53.18\%$ of App were voluntary. A $\chi^2$-test shows that differences are significant ($\chi^2(4, N=775)=14.68$, $p<0.001$). 
No participant reported using 2F for any motivation other than forced/incentive/voluntary (participants could do so via an open-ended question).

\begin{figure}
\centering
  \includegraphics[scale=.8]{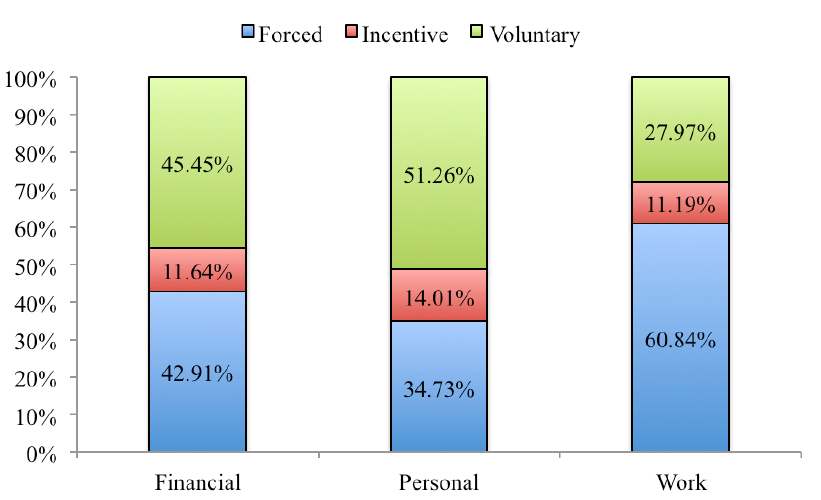}
  \caption{Distribution of the motivation across contexts.}
  \label{fig:motivationvscontext} 
\end{figure}

\descr{Different Motivations in Different Contexts.}
We find that in the work context, $60.84\%$ of participants were forced to use 2F, versus $27.97\%$ of participants using 2F voluntarily. In the personal context, more than half participants ($51.26\%$) use 2F voluntarily and $34.73\%$ are forced to. In the financial context, about $45.45\%$ of participants use 2F voluntarily and $42.91\%$ are forced to. Distributions
are plotted in Figure~\ref{fig:motivationvscontext}. A $\chi^2$-test shows that differences are significant ($\chi^2(4, N=775)=29.76$, $p<0.0001$).

This result is expected, as users tend to be forced to use 2F at work, and tend to use it voluntarily (opt-in) for personal use. In the financial context, the distribution is even. 

\begin{table}[ttt]
\begin{center}
\small
\begin{tabular}{| c | c | c |}
  \hline                        
   & \textbf{Female} & \textbf{Male} \\
  \hline    
  App Users & 31 & 71 \\\hline  
  Non-App & 53 & 64 \\\hline  
\end{tabular}
\end{center}
\vspace{0.1cm}
\caption{Distribution of gender across 2F App technology.}\label{tab:gender}
 \end{table}
 
\descr{Gender Differences.}
While there is no gender difference in terms of adoption rate for Token and Email/SMS, male users adopt App-based 2F more than female users -- see Table~\ref{tab:gender}. The $\chi^2$-test shows that the difference is significant ($\chi^2(1, N=219)=29.76$, $p<0.05$).

\descr{2F for Online Gaming.}
As mentioned earlier, we also asked participants to list services and websites for which they used 2F. Surprisingly, we find that in the personal context, in addition to personal email, document sharing, and social networking sites, participants also used 2F for online gaming, e.g., on Battle.net, Diablo 3, World of Warcraft, Blizzard Entertainment, and swtor.com.

\subsection{Exploratory Factor Analysis}

\begin{table*}[ttt]
\begin{center}
\small
\begin{tabular}{| c | | c | c | c | c |}
  \hline                    
   &    \multicolumn{4}{c}{\textbf{Loadings}} \\
  \hline                        
   & \textbf{Factor 1: Ease of Use} & \textbf{Factor 2: Cognitive Efforts} & \textbf{Factor 3: Trust} & \textbf{Communality} \\
  \hline      \hline    
  \textsf{\small Convenient} & \textbf{0.91} & 0.05 & -0.02 & 0.77 \\ \hline  
  \textsf{\small Quick} & \textbf{0.84} & -0.12 & -0.15 & 0.67 \\ \hline  
  \textsf{\small Enjoy} & \textbf{0.77} & 0.15 & 0.12 & 0.63 \\ \hline  
  \textsf{\small Reuse} & \textbf{0.75} & 0.04 & 0.19 & 0.75 \\ \hline  
  \textsf{\small Helpful} & \textbf{0.72} & 0.02 & 0.17 & 0.69\\ \hline  
  \textsf{\small No Enjoy} & \textbf{-0.52} & 0.22 & -0.16 & 0.55 \\ \hline  
  \textsf{\small User Friendly} & \textbf{0.42} & -0.19 & 0.37 & 0.74 \\ \hline  
  \textsf{\small Need Instructions} & 0.15 & \textbf{0.80} & -0.12 & 0.60\\ \hline  
  \textsf{\small Concentrate} & 0.03 & \textbf{0.64} & 0.14 & 0.38 \\ \hline  
  \textsf{\small Stressful} & -0.41 & \textbf{0.51} & 0.01 & 0.59 \\ \hline  
  \textsf{\small Match} & -0.30 & \textbf{0.42} & -0.15 & 0.47 \\ \hline  
  \textsf{\small Frustrating} & -0.47 & \textbf{0.47} & 0.00 & 0.63 \\ \hline  
  \textsf{\small Trust} & 0.08  & -0.04  & \textbf{0.80} & 0.74\\ \hline      
  \textsf{\small Secure} & -0.02 & 0.03 & \textbf{0.82} & 0.82\\ \hline      
  \textsf{\small Easy} & 0.27 & -0.28 & 0.31 & 0.44 \\ \hline \hline        
  Eigenvalues & 7.52 & 1.78 & 1.03 & \\ \hline 
  \% of Variance & 32 & 15 & 14 &  \\ \hline 
  Total Variance & & 61\% & &\\ \hline 
\end{tabular}
\end{center}
\vspace{0.2cm}
\caption{Factor Analysis Table.}\label{tab:factor}
 \end{table*}

While SUS is a generic usability measure, we argue that 2F technologies rely on a unique combination of hardware and software that SUS may fail to capture. Following this shortcoming, previous work~\cite{bonneau2012quest,braz2006security,gunson2011user,karole2011comparative,weir2009user,weir2010usable} considered a series of questions and parameters to evaluate the usability of  2F schemes. In order to obtain key elements central to the understanding of the usability of 2F, we perform an exploratory factor analysis (see aforementioned $15$ Likert questions). 

We factor-analyze our questions using Principal Component Analysis (PCA) with Varimax (orthogonal) rotation. Items with loadings $< 0.4$ are excluded. The analysis yields three factors explaining a total of $61\%$ of the variance for the entire set of variables. These factors are independent of each other (i.e., they are not correlated).

Factor $1$ is labeled \textbf{\em Ease of Use} (EaseUse for short) due to the high loadings by following items: \textsf{\small Quick}, \textsf{\small enjoy}, \textsf{\small user friendly}, \textsf{\small convenient}, \textsf{\small easy reuse}, \textsf{\small helpful}, and \textsf{\small convenient}. This first factor explains $32\%$ of the variance. 

The second derived factor is labeled \textbf{\em Cognitive Efforts} (CogEfforts for short). This factor is labeled as such due to the high loadings by following factors: \textsf{\small Frustrating}, \textsf{\small stressful}, \textsf{\small match}, \textsf{\small need instructions}, and \textsf{\small concentration}. The variance explained by this factor is $15\%$. 

The third derived factor is labeled \textbf{\em Trustworthiness}. This factor is labeled as such due to the high loadings by following factors: \textsf{\small Secure} and \textsf{\small trust}. The variance explained by this factor is $14\%$ (Table~\ref{tab:factor}).

The commonalities of the variables included are rather low overall, with one variable having a small amount of variance (\textsf{\small Concentrate}, $38\%$) in common with the other variables in the analysis. This may indicate that the variables chosen for this analysis are only weakly related with each other. However, the KMO test indicates that the set of variables are at least adequately related for factor analysis.

In conclusion, we have identified three clear factors among participants: ease of use, required cognitive efforts, and trustworthiness.

\begin{figure}[ttt]
\centering
  \includegraphics[width=0.7\linewidth]{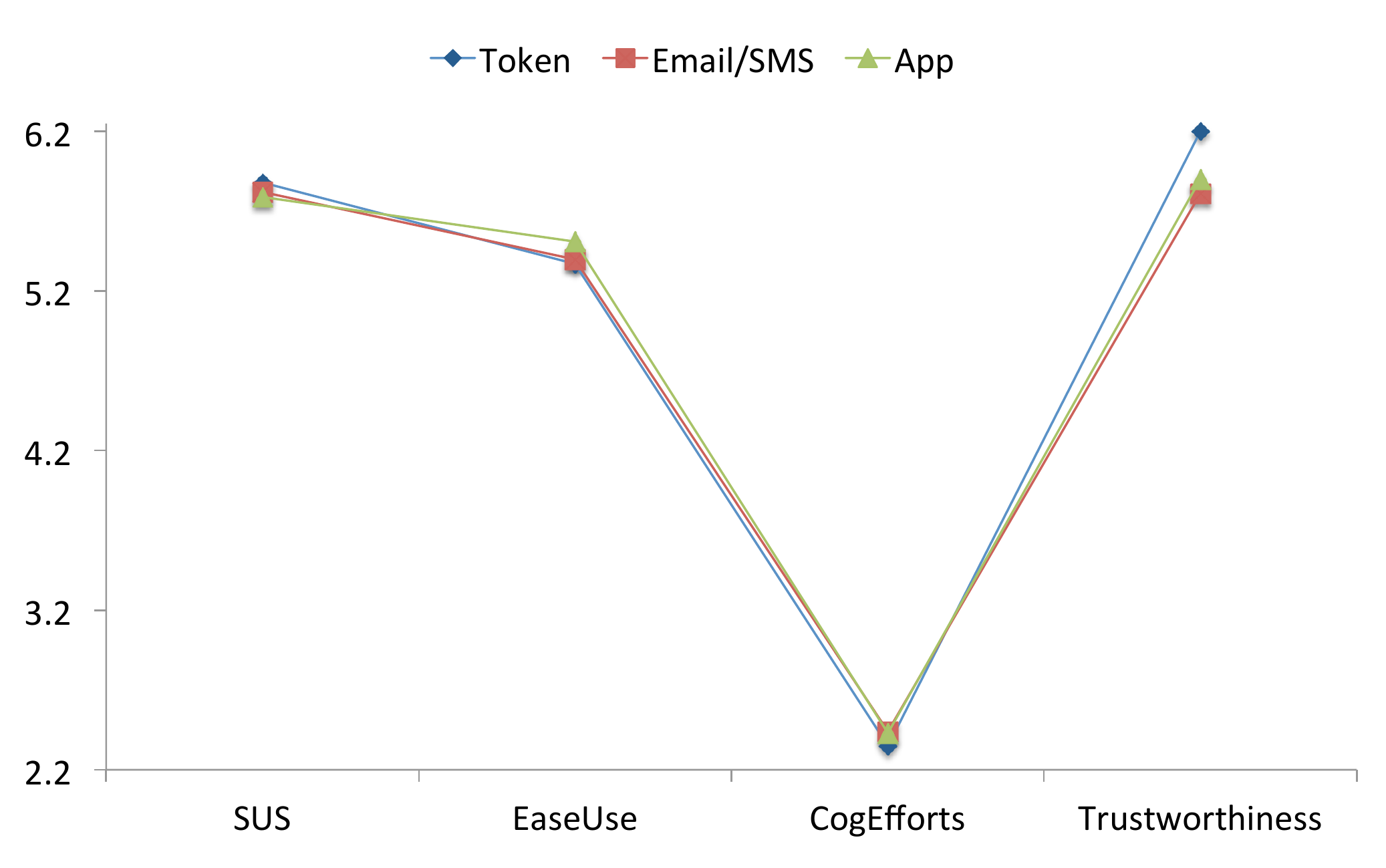}
\caption{Overview of usability measures of different 2F technologies. The x-axis lists the different considered factors and the y-axis gives the average score on the 7-point Likert scale. }
  \label{fig:overview} 
\end{figure}

\descr{Overview of Usability Measures.} In Figure~\ref{fig:overview}, we show the average usability measures for different 2F technologies. We obtain similar ratings for different technologies. The average SUS score is around 5.8, EaseUse is about 5.4, CogEfforts is 2.4, and Trustworthiness is around 6. We observe that SUS is high for all 2F technologies: converting the SUS score to a percentage scale, overall SUS is more than $80\%$, which is considered ``Grade A'' usability~\cite{sauro2012quantifying}. In addition, SUS is correlated with EaseUse ($r=0.8$). 
Next, we look into factors that influence 2F usability measures. 

\descr{Comparison among Different 2F Technologies.} We now compare the usability of different 2F technologies, taking into consideration the context in which they are used and the characteristics of the individuals who used them, including age, gender and whether they have a computer science background (``CS\_Back''). Some participants only use one of the 2F technologies and others use more than one. To compare the usability of different technologies, we split participants into $7$ subgroups (Table~\ref{tab:usage}) and performed analysis on each of the subgroups. Since there are not enough participants in Groups $1$, $3$, and $5$~(Table~\ref{tab:usage}), these groups were not analyzed. For usability measures, we use the three newly  discovered factors (introduced above): ``EaseUse'' ($\alpha = 0.92$), ``CogEffort'' ($\alpha = 0.74$), and ``Trustworthiness'' ($\alpha = 0.81$).

\descr{Email/SMS Users.}
77 participants only use Email/SMS as 2F technology, with 13 of them having a CS background. To compare the usability measures for participants who only use Emails/SMS (Group 2), we ran a MANOVA with one between factor computer science background (CS\_Back vs non CS\_Back), and age, gender and context as covariates. The dependent variables were the three usability measures. Using Pillai's trace, CS\_Back  ($V= 0.07, F(3,124)=3.04, p=0.02$) was a significant factor. Age, gender and context were not significant.

We conduct further analysis to test the effects of computer science background on the 3 usability dimensions using Mann-Whitney U Test (i.e., Shapiro-Wilk test shows that our data is not normally distributed). We used Bonferroni adjusted alpha levels of $0.0167$ per test ($0.05/3$). Results indicate that participants without a computer science background (EaseUse $Md= 5.88$) find Email/SMS to be easier to use than participants with a computer science background (EaseUse $Md = 4.88$, $U = 1792$, $p = 0.001$).

\descr{Token \& Email/SMS Users.}
29 participants used both token and email/SMS 2F technologies, with 7 having a computer science background. To compare the usability measures for participants who use both Emails/SMS and app (Group 4), we ran a one way (Technology: Token vs. Email/SMS) within subject MANOVA, with age, gender and context as covariates. The CS\_Back was not included in the analysis because there were not enough participants with CS\_Back. No main effect of technology was found. Age was a significant covariate ($V= 0.13, F(3,63)=3.12, p=0.03$).

Similarly, we test the effects of age on the 3 usability dimensions using Mann-Whitney U Test  and Bonferroni adjusted alpha levels of $0.0167$ per test ($0.05/3$). Results show that elder people in Group 4 (age above median age of $33; N = 12$, cogEfforts $Md = 3$) need more cognitive efforts to use 2F technology than younger people ($N=17$, CogEfforts $Md = 2, U = 873, p = 0.003$).

\descr{Email/SMS \& App Users.} 50 participants used both Email/SMS and App (Group 6), with only 5 having a computer science background. To compare the usability measures for participants who use both Emails/SMS and app (Group 6), we ran a one way (Technology: Email/SMS vs. App) within subject MANOVA, with age, gender and context as covariates. The CS\_Back was not included in the MANOVA because there were not enough participants. Similar to results in Group 4, no main effect of technology was found. Age was a significant covariate ($V= 0.13, F(3,63)=3.12, p=0.03$).

Similarly, we test the effects of age on the 3 usability dimensions using Mann-Whitney U Test  and Bonferroni adjusted alpha levels of $0.0167$ per test ($0.05/3$). Results show that elder people find 2F technology less trustworthy (Trustworthiness $Md = 5.5$) than younger people (Trustworthiness $Md=6.0, U=2755, p =0. 007$).

\descr{Token, Email/SMS \& App Users.} 41 participants used all token, email/SMS and App, with 17 having a CS background. To compare the usability measures for participants in this group, we ran a 3(Technology: Token, Email/SMS vs. App) x 2(CS\_Back vs non CS\_Back) MANOVA, with Technology as a within subject variable and CS\_Back as a between subject variable, and age, gender and context as covariates. Technology and CS\_back were not significant. Gender is a significant factor ($V= 0.12, F(3,168) = 7.44, p = 0.0001$).

Similarly, we test the effects of gender on the 3 usability dimensions using Mann-Whitney U Test  and Bonferroni adjusted alpha levels of $0.0167$ per test ($0.05/3$). Female users (CogEfforts $Md = 2.75, N = 12$) need more cognitive efforts than male users (CogEfforts $Md = 2.00, U= 4124, p = 0.001, N = 29$).

\subsection{Analysis of Open-Ended Questions}
For each 2F technology, we asked users to answer a few open-ended questions about the services/websites where the used 2F and they issues they encountered. 
As mentioned earlier, security tokens tend to be used for work, finance, and personal websites. Interestingly, users often rely on tokens to protect their online gaming accounts: the fear of losing their gaming profile is high enough for users to adopt 2F. Users complain that the authentication process is often prone to failure (``The authentication to the server was down.''), is time sensitive (``Sometimes, during the code rollover, you'd end up with a mismatch and have to start the whole process over''), and that problem resolution is complicated (``If I made three mistakes entering my code, I had to call the state help desk to have my PIN reset''). 

Email/SMS have, overall, a high variety of use cases, but were frequently used with banks as well as with Facebook, Google, and Paypal. People complained about specific issues with codes expiring or failing to be received, especially while traveling abroad. For instance, a number of users complain about SMS not working abroad (``Sometimes it wouldn't send'', ``My husband changed his phone number when moving to the US and had a lot of problems getting things.'', ``Sometimes I am unable to receive a code if I am overseas. In that case, I have to call a toll free number or e-mail customer support to receive the code via e-mail instead of text.''), and again regarding the difficult problem resolution (``The passcode they sent me didn't work and I had to call them to get a new one. It was very frustrating.''). 

Finally, we noticed that enterprises rely on (mostly proprietary) security tokens (e.g., RSA/Verisign tokens) for authentication to corporate networks in the workplace. Also, smartphone apps (e.g., Google Authenticator) are mostly used by customers who opt-in to 2F with online services providers, such as, Google, Dropbox, or Facebook.

\section{Discussion}\label{sec:discussion}
We now discuss findings drawn by our exploratory study, and highlight items for future work.

\descr{Adoption.} 2F technologies are adopted at different rates, depending on {\em contexts} and {\em motivations}. Specifically, in the work environment, codes generated by security tokens constitute the most used second factor of authentication. Codes received via email or SMS are most popular in the financial and personal contexts. Also, few users receive incentives to adopt 2F, while many utilize security tokens because they are forced to, or decide to opt-in to use dedicated smartphone apps. 

\descr{Usability.} User perceptions of the usability of 2F is often correlated with their individual characteristics (e.g., age, gender, background), rather than with the actual technology or the context/motivation in which it is used. We find that, overall, 2F technologies are perceived as highly usable, with little difference among them, not even when they are used for different motivations and different contexts. This seems to  contrast with prior work on password policies~\cite{inglesant2010true}
which showed that context of use impacts the ability of users to become 
familiar with complex passwords and, ultimately, affects their usability.

One possible explanation, supported by participants' responses to open-ended questions, is that 
most 2F users do not need to provide the second authentication factor very often. For instance, some financial institutions (e.g., Chase and Bank of America) or services provider (such as Google and Facebook) only require the second factor to be entered if a user is authenticating from an unrecognized device (e.g., from a new location or after clearing cookies).

\descr{Trustworthiness.} Another relevant finding is that users' perception of trustworthiness is not negatively correlated with ease of use and required cognitive efforts, somewhat in contrast to prior work~\cite{braz2006security,gunson2011user}. We find that 2F technologies perceived as more trustworthy are not necessarily less usable. One possible explanation is that prior work mostly compared 2F with passwords.

\descr{Impact.} We argue that our comparative analysis is essential to begin assessing attitudes and perceptions of 2F users, identifying causes of friction, driving user-centered design of usable 2F technologies, as well as informing future usable security research.
Note that, in many cases, authentication based on passwords only is actually not an option (e.g., for corporate VPN access, or for some financial services), and thus more usable 2F technologies in that context should be favored 
to avoid friction~\cite{adams1999users}, negative impact on productivity~\cite{strouble2009productivity,inglesant2010true}, as well as driving users to circumvent authentication policies they perceived as unnecessarily stringent~\cite{inglesant2010true}.
Similarly, when users have the choice to opt-in, adoption rates will likely depend on 2F usability.

\descr{Limitations and future work.} We acknowledge that our work presents some limitations and leaves a few items to future work. First, it is based on a survey of 219 MTurk users, who, arguably, might be more computer savvy and might adopt 2F more than the general population. Second, some of the points raised by our analysis -- such as, non-correlation of usability and context/motivation of use as well as the usability metrics derived by our factor analysis -- should be validated by open-ended interviews and qualitative studies. Indeed, our current and future work includes the design of a real-world user study building on the experience and the findings from this work.

\section{Conclusions}
This paper presented an exploratory comparative study of two-factor authentication (2F) technologies. First, we reported on a pre-study interview involving 9 participants, intended to identify popular 2F technologies as well as how they are used, when, where, and why. Next, we designed and administered an online survey to 219 Mechanical Turk users, aiming to measure the usability of a few popular 2F technologies: one-time codes generated by security tokens, one-time PINs received via SMS or email, and dedicated smartphone apps. We also recorded contexts and motivations, and study their impact on perceived usability of different 2F technologies. We considered participants that used specific 2F technologies, either because they were forced to, or because they wanted to.

We presented an exploratory factor analysis to evaluate a series of parameters, including some suggested by previous work, to evaluate the usability of 2F, and show that ease of use, trustworthiness, and required cognitive effort are the three key aspects defining 2F usability. Finally, we showed that differences among the usage of 2F depend on individual characteristics of people, more than the actual technologies or contexts of use. We considered a few characteristics, such as age, gender and computer science background, and obtained a few insights into user preferences. 

Our preliminary study is essential to guide and inform the design of follow-up qualitative studies, which we plan to conduct as part of future work.

\balance
\bibliographystyle{abbrv}
\bibliography{bibfile}

\end{document}